\documentstyle[aps,amsfonts,prd]{revtex}

\begin{document}

\title {Resonant photon creation in a three dimensional oscillating
cavity}
\author{Martin Crocce $^{1}$ \thanks{mcrocce@df.uba.ar}, 
Diego A.\ R.\ Dalvit $^{2}$ \thanks{dalvit@lanl.gov},
and Francisco D.\ Mazzitelli $^{1}$ \thanks{fmazzi@df.uba.ar} }
\address{$^1$ 
Departamento de F\'\i sica {\it J.J. Giambiagi}, 
Facultad de Ciencias Exactas y Naturales\\ 
Universidad de Buenos Aires - Ciudad Universitaria, Pabell\' on I, 
1428 Buenos Aires, Argentina} 
\address{$^2$ T-6, Theoretical Division, MS B288, Los Alamos National
Laboratory, Los Alamos, NM 87545}

\maketitle

\begin{abstract}
We analyze the problem of photon creation inside a perfectly conducting, 
rectangular, three dimensional cavity
with one oscillating wall. For some particular values of
the frequency of the oscillations
the system is resonant. We solve the field equation using multiple
scale analysis and show that the total number of photons inside
the cavity grows exponentially in time. 
This is also the case for slightly off-resonance situations.
Although the spectrum of
a cavity is in general non equidistant, we show that 
the modes of the electromagnetic field can be coupled, and that 
the rate of photon creation
strongly depends on this coupling. We also analyze    
the thermal enhancement of the photon creation.

\end{abstract}

\pacs{PACS number(s): 42.50.Lc, 42.50.Dv, 12.20.-m}


\section{Introduction}

The existence of an attractive force between two perfectly conducting 
plates has been predicted by Casimir in 1948 
\cite{Casimir}. It has been measured with accurate precision in the 
last years by  Lamoreaux \cite{Lamoreaux} and Mohideen \textit{et al.}
\cite{Mohideen}. These experiments confirm the existence of vacuum 
field fluctuations in the framework of field quantization with static 
boundaries, and increase the interest in the case of dynamical 
boundaries as well.

The dynamical effect consists in the generation of photons due to the 
instability of the vacuum state of the electromagnetic field 
in the presence of time-dependent boundaries. In the 
literature it is referred to as dynamical Casimir effect \cite{Schwinger} 
or motion-induced radiation \cite{Lambrecht}.
Up to now no concrete experiment has been carried 
to confirm this photon generation, but an experimental 
verification is not out of reach. 
From the theoretical point of view it is widely accepted that the most 
favorable configuration in order to observe the phenomenon is a 
vibrating cavity in which it is possible to  produce resonant effects 
between the mechanical and field oscillations.

Many previous papers have focused their attention in the field quantization 
within a one dimensional cavity with one or two  walls performing small 
amplitude oscillations, 
at twice the eigenfrequency of some unperturbed electromagnetic mode.
For these cavities there 
exists a strong intermode interaction, which is a consequence of the 
equidistant character of the frequencies spectrum.
The main features of the one dimensional model are that photons are 
created in all electromagnetic modes (due to mode-mode coupling), 
and that the total energy inside the cavity grows exponentially 
at the expense of the energy given to the system to keep the wall moving.
A simple approach is to make a perturbative expansion in terms of the 
amplitude of oscillations, as was done in \cite{Ji}. However this 
perturbative treatment breaks down after a short period of time due to 
the appearence of secular terms.
In \cite{Mazzitelli} the renormalization group technique was used in 
order to obtain a solution valid for a period of time longer than that 
of the pertubative case. There it was shown that the energy spectrum 
develops a non trivial structure formed by peaks travelling at the speed of 
light and bouncing against the walls (in agreement with other authors 
\cite{Cole}).
Reference \cite{Dodonov} makes use of the fact that two different 
time scales characterize the problem; the usual one, related to the 
wall's oscillation period, and a `slow' one which accounts for the 
cumulative resonance effect. It is then posible to isolate the resonant 
part after averaging over fast oscillations the initial equations for the 
electromagnetic field modes.

There are some works in the literature dealing with higher 
dimensional cavities . In \cite{Neto}  
the radiation emitted in each polarization of the electromagnetic field
was computed perturbatively, 
when two parallel, plane, and perfectly conducting plates oscillate  
along the direction perpendicular to their 
surfaces. Such geometry constitutes the simplest example 
of an open three dimensional cavity.
In \cite{Soh} the authors obtained  the  distribution of the created 
photons for the case of parametric resonance inside a three dimensional cavity.
In both cases \cite{Neto,Soh}
a perturbative method was applied so the results  
are valid in the short time limit.
In \cite{Dodonov} a nonperturbative analysis was presented, generalizing the 
method of averaging over fast oscillations to higher-dimensional cavities. 
However, the intermode coupling was neglected, reducing the problem to that of 
one single parametric oscillator.

Of particular interest is to find out how the finite temperature affects 
the photon production. This was studied in \cite{Plunien} with a 
nonperturbative method (see also \cite{LR}). 
A remarkable enhancement of the pure vacuum effect was 
found, but again neglecting the coupling between modes. 

In this paper we present  a detailed analysis of the photon generation 
inside a three dimensional resonant cavity. We also discuss 
the finite temperature case,  showing the enhancement of 
the effect with respect to $T=0$. We apply the Multiple Scale Analysis (MSA) 
which provides us with a simple 
technique equivalent to summing the most secular terms to all orders in the 
perturbative treatment. In this way 
we can get a solution valid for a period of 
time longer than that of the perturbative case.
We pay particular attention to the resonant coupling  between 
different modes.

The paper is organized as follows. In section II we obtain the time 
evolution of the quantized field by expanding it over the 
\textit{`instantaneous basis'}. For simplicity we  deal with a 
scalar bosonic field. Following the steps given in \cite{Law} we arrive 
at an infinite set of coupled differential equations for the coefficients 
of the expansion. We also explain there how to compute the number of 
motion-induced photons for the zero temperature case. In section III we 
describe and apply the MSA to our problem. We find the  
coupling conditions between different modes that can be satisfied 
depending on the cavity's spectrum. In section IV we present  a 
general analysis of the coupling conditions, and discuss some examples.
In particular we find that the fundamental mode of a cubic cavity is coupled 
to another mode in the parametric resonance case, giving as result that the 
number of  photons with two different frequencies increases 
exponentially in time. However, the production rate for the fundamental mode 
is only one half of that expected if we had neglected the coupling,
as previous works did. At the end of this section we study slightly
off-resonance situations. 
In section V we obtain an expression for the number of 
photons in each mode assuming that the field was initially in thermal 
equilibrium. Section VI contains our final remarks 
and comments on the
generalization to the more realistic case of the electromagnetic field.


\section{scalar field quantization with moving boundaries}

We consider a rectangular cavity formed by perfectly reflecting walls 
with dimensions $L_{x},L_{y},$ and $L_{z}$. The wall placed at $x=L_{x}$ 
is at rest for $t<0$ and begins to move following a given trayectory, 
$L_{x}(t)$, at $t=0$. Note that we assume this trayectory as prescribed for 
the problem (not a dynamical variable) and that it works as a time-dependent 
boundary condition for the field.
The field $\phi(\mathbf{x},\mathit{t})$ satisfies the wave equation 
$\Box\phi=0$  in $3+1$ dimensions, and the boundary conditions 
$\phi|_{\rm walls}=0$ for all times.
The Fourier expansion of the field for an arbitrary moment of time, 
in terms of creation and annihilation operators, can be written as

\begin{equation}
\phi(\mathbf{x},\mathit{t})=\sum_{\mathbf{n}}
\hat{a}_{\mathbf{n}}^{\scriptscriptstyle{\rm in}}
u_{\mathbf{n}}(\mathbf{x},\mathit{t}) + {\rm H.c.} ,
\label{field}
\end{equation} 
where the mode functions $u_{\mathbf{n}}(\mathbf{x},\mathit{t})$ form
a complete orthonormal \footnote{The inner product is the usual for 
Klein-Gordon equation, namely
$ (\psi,\xi)=-i\int_{\rm cavity}d^{3}\!x\,[\,\psi\,\dot{\xi^{\star}}-
\dot{\psi}\,\xi^{\star}\,] . $} set of solutions of the wave equation 
with vanishing boundary conditions. 

When $t\leq0$ (static cavity) each field mode is 
determined by three positive integers $n_{x},n_{y}$ and $n_{z}$. Namely

\begin{equation}
u_{\mathbf{n}}(\mathbf{x},\mathit{t}<0)={1\over\sqrt{2\omega_
{\mathbf{n}}}}\sqrt{\frac{2}{L_{x}}}\sin\left(\frac{n_{x}\pi}
{L_{x}} x\right)\sqrt{\frac{2}{L_{y}}}
\sin\left(\frac{n_{y}\pi}{L_{y}} y\right)
\sqrt{\frac{2}{L_{z}}}\sin\left(\frac{n_{z}\pi}{L_{z}} z\right)
e^{i\omega_{\mathbf{k}}t} ,
\label{expest}
\end{equation}
\begin{equation}
\omega_{\mathbf{n}}=\pi\sqrt{\left(\frac{n_{x}}{L_{x}}\right)^{2} \!+ 
\left(\frac{n_{y}}{L_{y}}\right)^{2}\! + 
\left(\frac{n_{z}}{L_{z}}\right)^{2}} ,
\label{omega}
\end{equation}
with the shorthand $\mathbf{n}=(\mathit{n}_{x},\mathit{n}_{y},\mathit{n}_{z})$
\footnote{We are using units $\hbar=c=1$.}.

When $t>0$ the boundary condition on the moving wall becomes 
$\phi(x=L_{x}(t),y,z,t)=0$. In order to satisfy it we expand the mode 
functions in Eq.(\ref{field}) with respect to an \textit{instantaneous basis}  

\begin{eqnarray}
u_{\mathbf{n}}(\mathbf{x},\mathit{t}>0)&=&\sum_{\mathbf{k}}
Q_{\mathbf{k}}^{(\mathbf{n})}(t)\sqrt{\frac{2}{L_{x}(t)}}
\sin\left(\frac{k_{x}\pi}{L_{x}(t)} x\right)
\sqrt{\frac{2}{L_{y}}}\sin\left(\frac{k_{y}\pi}{L_{y}} y\right)
\sqrt{\frac{2}{L_{z}}}\sin\left(\frac{k_{z}\pi}{L_{z}} z\right)\\
&=&\sum_{\mathbf{k}}Q_{\mathbf{k}}^{(\mathbf{n})}(t)\,
\varphi_{\mathbf{k}}(\mathbf{x},\mathit{L}_{x}(t)) ,
\label{exp}
\end{eqnarray}
with the initial conditions

\begin{equation}
Q_{\mathbf{k}}^{(\mathbf{n})}(0)={1\over\sqrt{2\omega_{\mathbf{n}}}}\,
\delta_{\mathbf{k},\mathbf{n}}  \ ,\ \ 
\dot{Q}_{\mathbf{k}}^{(\mathbf{n})}(0)=-i
\sqrt{\omega_{\mathbf{n}}\over2}\,\delta_{\mathbf{k},\mathbf{n}} .
\end{equation}
In this way we ensure that, as long as $L_{x}(t)$ and 
$\dot{L}_{x}(t)$ are continuous at $t=0$,
each field mode and its time derivate are also continuous functions. 
The expansion in Eq.(\ref{exp}) for the field modes must be a solution of the 
wave equation. Taking into account that the $\varphi_{\mathbf{k}}$'s 
form a complete and orthonormal set and that they depend on $t$ only 
through $L_{x}(t)$, we obtain a set of coupled 
equations for $Q_{\mathbf{k}}^{\mathbf{n}}(t)$\cite{Law}:
\begin{equation}
\ddot{Q}_{\mathbf{k}}^{(\mathbf{n})}+\omega_{\mathbf{k}}^{2}(t)\,
Q_{\mathbf{k}}^{(\mathbf{n})}=2\lambda(t)\sum_{\mathbf{j}}
g_{\mathbf{kj}}\,\dot{Q}_{\mathbf{j}}^{(\mathbf{n})}+\dot{\lambda}(t)
\sum_{\mathbf{j}}g_{\mathbf{kj}}\,Q_{\mathbf{j}}^{(\mathbf{n})}+
\lambda^{2}(t)\sum_{\mathbf{j,l}}g_{\mathbf{lk}}\,g_{\mathbf{lj}}\,
Q_{\mathbf{j}}^{(\mathbf{n})} ,
\label{ecacop}
\end{equation}
where
\begin{equation}
\omega_{\mathbf{k}}(t)=\pi\sqrt{\left(\frac{k_{x}}{L_{x}(t)}\right)^{2} \,+ 
\left(\frac{k_{y}}{L_{y}}\right)^{2}\, + \left(\frac{k_{z}}{L_{z}}\right)^{2}}
\ \ ; \ \ \lambda(t)=\frac{\dot{L}_{x}(t)}{L_{x}(t)} .
\end{equation}
The coefficients $g_{\mathbf{kj}}$ are defined by
\begin{equation}
g_{\mathbf{kj}}=L_{x}(t)\int_{0}^{L_{x}(t)}dx\ 
\frac{\partial\varphi_{\mathbf{k}}}{\partial L_{x}}\,\varphi_{\mathbf{j}},
\end{equation}
and read
\begin{equation}
g_{\mathbf{kj}}=-g_{\mathbf{jk}}= \left\{ 
\begin{array}{ll}
(-1)^{k_{x}+j_{x}}\frac{2k_{x}j_{x}}{j_{x}^{2}-k_{x}^{2}}\,
\delta_{k_{y}j_{y}}\,\delta_{k_{z}j_{z}} \,& \mbox{if $k_{x}\neq j_{x}$} \\ 
0 & \mbox{if $k_{x}=j_{x}$}  .
\end{array}
\right.
\end{equation}
Furthermore, in deriving Eq.(\ref{ecacop}) we have used that
$\sum_{\mathbf{l}}\,g_{\mathbf{kl}}\,g_{\mathbf{jl}}= 
L_{x}^{2}\int dx\, (\partial\varphi_{\mathbf{k}}/\partial L_{x})
(\partial\varphi_{\mathbf{j}}/\partial L_{x})$, 
which follows from the completeness relation of the $\varphi_{\mathbf{k}}$'s.

The annihilation and creation operators $\hat{a}_{\mathbf{k}}^
{\scriptscriptstyle{\rm in}}$ and $\hat{a}^{\dag\,\scriptscriptstyle{\rm in}}_
{\mathbf{k}}$ correspond to the particle notion in the `in' 
region ($t<0$). If the wall stops for $t>t_{\rm final}$, we can define a new 
set  of operators, $\hat{a}_{\mathbf{k}}^{\scriptscriptstyle{\rm out}}$ and 
$\hat{a}^{\dag\,\scriptscriptstyle{\rm out}}_{\mathbf{k}}$, associated to 
the particle notion in the `out' region ($t>t_{\rm final}$). 
These two sets of operators are connected by means of the Bogoliubov 
transformation
\begin{equation}
\hat{a}_{\mathbf{k}}^{\scriptscriptstyle{\rm out}}=
\sum_{\mathbf{n}} ( \hat{a}_{\mathbf{n}}^{\scriptscriptstyle{\rm in}}\,
\alpha_{\mathbf{nk}}+\hat{a}^{\dag\,
\scriptscriptstyle{\rm in}}_{\mathbf{n}}\,\beta_{\mathbf{nk}}^{\star} ) .
\label{bog1}
\end{equation}
The coefficients $\alpha_{\mathbf{nk}}$ and $\beta_{\mathbf{nk}}$ 
can be obtained as follows. When the wall returns to its initial position 
the right hand side in Eq.(\ref{ecacop}) vanishes and the solution reads
\begin{equation}
Q_{\mathbf{k}}^{(\mathbf{n})}(t>t_{\rm final})=
A_{\mathbf{k}}^{(\mathbf{n})}e^{i\omega_{\mathbf{k}}t}+
B_{\mathbf{k}}^{(\mathbf{n})}e^{-i\omega_{\mathbf{k}}t},
\label{sol}
\end{equation}
with $A_{\mathbf{k}}^{(\mathbf{n})}$ and $B_{\mathbf{k}}^{(\mathbf{n})}$ 
being some constant coefficients to be determined by the continuity 
conditions at $t=t_{\rm final}$. 
Inserting Eq.(\ref{sol}) into Eqs.(\ref{field}) and (\ref{exp}) we obtain 
an expansion of $\phi$ in terms of 
$\hat{a}_{\mathbf{k}}^{\scriptscriptstyle{\rm in}}$ and 
$\hat{a}_{\mathbf{k}}^{\dag\,\scriptscriptstyle{\rm in}}$ for 
$t>t_{\rm final}$. Comparing this with the equivalent expansion in terms of 
$\hat{a}_{\mathbf{k}}^{\scriptscriptstyle{\rm out}}$ and 
$\hat{a}_{\mathbf{k}}
^{\dag\,\scriptscriptstyle{\rm out}}$ it is easy to see that
\begin{equation}
\alpha_{\mathbf{nk}}=\sqrt{2\omega_{\mathbf{k}}} 
B_{\mathbf{k}}^{(\mathbf{n})}\ \ , \ \ \beta_{\mathbf{nk}}=
\sqrt{2\omega_{\mathbf{k}}}\,A_{\mathbf{k}}^{(\mathbf{n})} .
\label{bog2}
\end{equation}

The amount of photons created in the mode $\mathbf{k}$ is the average value 
of the number operator 
$\hat{a}_{\mathbf{k}}^{\dag\,\scriptscriptstyle{\rm out}}
\hat{a}_{\mathbf{k}}^{\scriptscriptstyle{\rm out}}$ with respect to the 
initial vacuum state (defined through 
$\hat{a}_{\mathbf{k}}^{\scriptscriptstyle{\rm in}}
|0_{\scriptscriptstyle{\rm in}}\rangle=0$). With the help of 
Eq.(\ref{bog1}) and Eq.(\ref{bog2}) we get
\begin{equation}
\langle {\mathcal{N}}_{\mathbf{k}} \rangle=\langle 
0_{\scriptscriptstyle{\rm in}}\mid 
\hat{a}_{\mathbf{k}}^{\dag\,\scriptscriptstyle{\rm out}} 
\hat{a}_{\mathbf{k}}^{\scriptscriptstyle{\rm out}}
\mid 0_{\scriptscriptstyle{\rm in}} \rangle = 
\sum_{\mathbf{n}}2\omega_{\mathbf{k}}|A_{\mathbf{k}}^{(\mathbf{n})}|^{2}
\label{numerodefotones} .
\end{equation}
 

\section{Multiple scale analysis}

Up to this point the equations are valid for an arbitrary motion of the wall 
(we only assume $L(0)=L_{0}$ and $\dot{L}(0)=0$ because the wall is at rest 
for $t<0$).
We are interested in the number of photons created inside the cavity, so it 
is natural to look for harmonic oscillations of the wall which could enhance 
that number by means of resonance effects for some specific external 
frequencies. So we study the following trayectory
\begin{equation}
L(t)=L_{0}(1+\epsilon\sin(\Omega t)+\epsilon f(t)) ,
\end{equation}
where $f(t)$ is some decaying function that allows us to meet the continuity 
conditions at $t=0$ (for example $f(t)=-\Omega t\, e^{-\alpha t}$). 
For small amplitudes of oscillations ($\epsilon \ll 1$), the equations for 
the modes Eq.(\ref{ecacop}) take the form
\begin{equation}
\ddot{Q}^{({\bf n})}_{\bf k} + \omega^2_{\bf k} Q^{({\bf n})}_{\bf k} =
2 \epsilon \left(\frac{\pi k_x}{L_x}\right)^2 \sin(\Omega t) 
Q^{({\bf n})}_{\bf k}
- \epsilon \Omega^2 \sin(\Omega t) \sum_{\bf j} g_{{\bf k}{\bf j}} 
Q^{({\bf n})}_{\bf j} 
+ 2 \epsilon \Omega \cos(\Omega t) \sum_{\bf j} g_{{\bf k}{\bf j}}
\dot{Q}^{({\bf n})}_{\bf j} + + \epsilon O(f) + O(\epsilon^2) ,
\label{eqqk}
\end{equation}
where $O(f)$ denotes terms proportional to $f$, $\dot f$ and $\ddot f$.

It is known that a naive perturbative solution of these equations in powers
of the displacement $\epsilon$ breaks down after a short amount of time, 
of order $(\epsilon\Omega)^{-1}$. 
This happens for those particular values of the external
frequency $\Omega$ such that there is
a resonant coupling with the eigenfrequencies of the static 
cavity. In this situation, to find a solution valid for 
longer times (of order $\epsilon^{-2}\Omega^{-1}$)
we use the multiple scale analysis (MSA) technique \cite{bender}. We introduce
a second timescale $\tau=\epsilon t$ and expand $Q^{({\bf n})}_{\bf k}$ as 
follows (we shall content ourselves with first order MSA)
\begin{equation}
Q^{({\bf n})}_{\bf k}(t) = Q^{({\bf n})(0)}_{\bf k}(t,\tau) + \epsilon
Q^{({\bf n})(1)}_{\bf k}(t,\tau) + O(\epsilon^2) .
\end{equation}
The derivatives with respect to the timescale $t$ read
\begin{eqnarray}
\dot Q^{({\bf n})}_{\bf k} &=& \partial_t Q^{({\bf n})(0)}_{\bf k} +
\epsilon [ \partial_{\tau}  Q^{({\bf n})(0)}_{\bf k} +
\partial_t Q^{({\bf n})(1)}_{\bf k} ] , \nonumber \\
\ddot Q^{({\bf n})}_{\bf k} &=& \partial^2_t Q^{({\bf n})(0)}_{\bf k} +
\epsilon [ 2 \partial^2_{\tau t}  Q^{({\bf n})(0)}_{\bf k} +
\partial^2_t Q^{({\bf n})(1)}_{\bf k} ] .
\end{eqnarray}
The initial conditions are

\begin{eqnarray}
Q^{({\bf n})(0)}_{\bf k}(0) &=& \frac{1}{\sqrt{2 \omega_{\bf k}}} 
\delta_{{\bf n},{\bf k}} , \nonumber \\
\dot{Q}^{({\bf n})(0)}_{\bf k}(0) &=& -i \sqrt{\frac{\omega_{\bf k}}{2}} 
\delta_{{\bf n},{\bf k}} .
\end{eqnarray}
To zeroth order in $\epsilon$ we get the equation of an harmonic oscillator

\begin{equation}
Q^{({\bf n})(0)}_{\bf k} = A^{({\bf n})}_{\bf k}(\tau) e^{i \omega_{\bf k} t}
+ B^{({\bf n})}_{\bf k}(\tau) e^{-i \omega_{\bf k} t} ,
\label{ansqk}
\end{equation}
and using the initial conditions it follows that

\begin{eqnarray}
A^{({\bf n})}_{\bf k}(\tau=0) &=& 0 ,
\label{iniA}\\
B^{({\bf n})}_{\bf k}(\tau=0) &=& \frac{1}{\sqrt{2 w_{\bf k}}} 
\delta_{{\bf n},{\bf k}} .
\label{iniB}
\end{eqnarray}

To first order in $\epsilon$ we obtain
\footnote{It is not straightforward to compute the next order corrections
using MSA. The introduction of new time scales like $\tau_1 = \epsilon t,
\tau_2 = \epsilon^2 t$, etc, is in general not sufficient to determine
the second order solution unambiguosly \cite{bender}. The renormalization
group method \cite{chen} (which is equivalent to MSA to first order
in $\epsilon$) seems to be more appropiate to improve systematically
the result. In any case, the next order corrections will be very
small for $\epsilon^2 t\leq\Omega^{-1}$.}

\begin{eqnarray}
\partial^2_t Q^{({\bf n})(1)}_{\bf k} + \omega^2_{\bf k} 
Q^{({\bf n})(1)}_{\bf k}
&=&
-2 \partial^2_{\tau t} Q^{({\bf n})(0)}_{\bf k} +
2 \left(\frac{\pi k_x}{L_x}\right)^2 \sin(\Omega t) Q^{({\bf n})(0)}_{\bf k}
\nonumber \\
&& - \Omega^2 \sin(\Omega t) \sum_{{\bf j} \neq {\bf k}} g_{{\bf k}{\bf j}}
Q^{({\bf n})(0)}_{\bf j} + 2 \Omega \cos(\Omega t)
\sum_{{\bf j} \neq {\bf k}} g_{{\bf k}{\bf j}}
\partial_t Q^{({\bf n})(0)}_{\bf j} + O(f)\,\, .
\end{eqnarray}

The basic idea of MSA is to impose that any term on the right-hand-side of the
previous equation with a time dependency of the form 
$e^{\pm i \omega_{\bf k} t}$ must vanish. If not, these terms would be in
resonance with the left-hand-side term and secularities would appear. 
The terms contained in $O(f)$ are not relevant because they are 
exponentially suppressed, and do not produce secularities. 
After imposing that no term $e^{+ i \omega_{\bf k} t}$ appear, we get
\begin{eqnarray}
{d A^{({\bf n})}_{\bf k}\over d \tau} &=&
-\frac{\pi^2 k_x^2}{2 \omega_{\bf k} L_x^2} B^{({\bf n})}_{\bf k}
\delta(2 \omega_{\bf k} -\Omega) +
\sum_{\bf j} (-\omega_{\bf j} + \frac{\Omega}{2})
\delta(-\omega_{\bf k} - \omega_{\bf j} + \Omega) 
\frac{\Omega}{2 \omega_{\bf k}} g_{{\bf k}{\bf j}} 
B^{({\bf n})}_{\bf j} \nonumber \\
&& + \sum_{\bf j} \left[
(\omega_{\bf j} + \frac{\Omega}{2}) 
\delta(\omega_{\bf k} - \omega_{\bf j} - \Omega) +
(\omega_{\bf j} - \frac{\Omega}{2}) 
\delta(\omega_{\bf k} - \omega_{\bf j} + \Omega)
\right]
\frac{\Omega}{2 \omega_{\bf k}} g_{{\bf k}{\bf j}} A^{({\bf n})}_{\bf j} .
\label{ec1}
\end{eqnarray}
In a similar fashion, the fact that no secularities should arise from the
$e^{-i \omega_{\bf k} t}$ term leads to

\begin{eqnarray}
{d B^{({\bf n})}_{\bf k}\over d\tau} &=&
-\frac{\pi^2 k_x^2}{2 \omega_{\bf k} L_x^2} A^{({\bf n})}_{\bf k}
\delta(2 \omega_{\bf k} -\Omega) +
\sum_{\bf j} (-\omega_{\bf j} + \frac{\Omega}{2})
\delta(-\omega_{\bf k} - \omega_{\bf j} + \Omega) 
\frac{\Omega}{2 \omega_{\bf k}} g_{{\bf k}{\bf j}} 
A^{({\bf n})}_{\bf j} \nonumber \\
&& + \sum_{\bf j} 
\left[
(\omega_{\bf j} + \frac{\Omega}{2}) 
\delta(\omega_{\bf k} - \omega_{\bf j} - \Omega) +
(\omega_{\bf j} - \frac{\Omega}{2}) 
\delta(\omega_{\bf k} - \omega_{\bf j} + \Omega)
\right]
\frac{\Omega}{2 \omega_{\bf k}} g_{{\bf k}{\bf j}} B^{({\bf n})}_{\bf j} .
\label{ec2}
\end{eqnarray} 
The previous set of two equations are non trivial (i.e., lead to resonant 
behavior) if $\Omega=2 \omega_{\bf k} \label{cond0}$ (resonant condition). 
Moreover, there is intermode coupling between modes ${\bf j}$ and 
${\bf k}$ if any of the following conditions is satisfied

\begin{eqnarray}
\Omega&=&\omega_{\mathbf{k}}+\omega_{\mathbf{j}} , \label{cond1}\\
\Omega&=&\omega_{\mathbf{k}}-\omega_{\mathbf{j}} , \label{cond2}\\
\Omega&=&\omega_{\mathbf{j}}-\omega_{\mathbf{k}} . \label{cond3}
\end{eqnarray}

There is an alternative, equivalent way of deriving the equations of 
motion \cite{Dodonov}. 
For $\epsilon\ll 1$, it is natural to assume that
the solution of Eq.(\ref{eqqk}) is of the form
\begin{equation}
Q^{({\bf n})}_{\bf k}(t) = A^{({\bf n})}_{\bf k}(t) e^{i \omega_{\bf k} t}
+ B^{({\bf n})}_{\bf k}(t) e^{-i \omega_{\bf k} t} ,
\label{ansqk2}
\end{equation}
where the functions  $A^{({\bf n})}_{\bf k}$ and 
$B^{({\bf n})}_{\bf k}$ are slowly varying. In order to obtain
differential equations for them, we insert this ansatz
into Eq.(\ref{eqqk}) and  neglect second derivatives of 
$A^{({\bf n})}_{\bf k}$ and $B^{({\bf n})}_{\bf k}$. After multiplying the 
equation by $ e^{\pm i \omega_{\bf k} t}$
we average over the fast oscillations. The resulting equations 
coincide with Eqs.(\ref{ec1}) and (\ref{ec2}).

We derived the equations for 3+1 dimensions. It is very easy to
obtain the corresponding ones in 1+1 and 2+1 dimensions. In all
cases the resonant conditions are given by Eqs.(\ref{cond1})-(\ref{cond3})
above. The main
difference between the 1+1 case and higher dimensions is that
in 1+1 the eigenfrequencies $\omega_k$
are proportional to integers. The
spectrum is equidistant and therefore an infinite set of modes may be
coupled. 
For example, when the external frequency is $\Omega
=2 \omega_1$, the mode $k$ is coupled with the modes $k\pm 2$.
This has been extensively studied in the literature 
\cite{Mazzitelli,Dodonov,Soh,Dod2}. 
In what follows we will be concerned with cavities with
non equidistant spectrum.


\section{Resonant photon creation}
\label{rpc}
In this section we shall solve the coupled Eqs.(\ref{ec1}) 
and (\ref{ec2}). We will see that there are different kinds of 
solutions depending both on the wall's frequency and the spectrum of the 
static cavity. Note that the spectrum is related to the cavity's dimensions 
through Eq.(\ref{omega}).
In subsection \ref{general} we will present a general analysis
of the resonant conditions and the solutions.  
We will show some particular examples in subsection \ref{examples}.

\subsection{General analysis}
\label{general}

Let us consider the `parametric resonance case', in which the frequency of 
the wall is twice the frequency of some unperturbed mode, say 
$\Omega=2\omega_{\mathbf{k}}$. Under this condition we expect that the number 
of created photons in the mode ${\mathbf{k}}$ will grow exponentially in 
time due to resonance effects. 
In order to find $A_{\mathbf{k}}^{(\mathbf{n})}$ and 
$B_{\mathbf{k}}^{(\mathbf{n})}$ from Eq.(\ref{ec1}) and Eq.(\ref{ec2}) we 
have to analyze whether the coupling conditions 
$|\omega_{\mathbf{k}}\pm\omega_{\mathbf{j}}|=\Omega$ can be satisfied or not. 
If we set $\Omega=2\omega_{\mathbf{k}}$, the resonant mode $\mathbf{k}$ will 
be coupled to some other mode $\mathbf{j}$ only if 
$\omega_{\mathbf{j}}-\omega_{\mathbf{k}}=\Omega=2\omega_{\mathbf{k}}$. 
Clearly, the latter relation will be satisfied depending on the spectrum of 
the particular cavity under consideration.

First, let us assume that this condition is not fullfilled. In this case, 
the equations for $A_{\mathbf{k}}^{(\mathbf{n})}$ and 
$B_{\mathbf{k}}^{(\mathbf{n})}$ reduce to

\begin{eqnarray}
\frac{dA^{(\mathbf{n})}_{\mathbf{k}}}{d\tau} & = & 
\frac{-1}{2\omega_{\mathbf{k}}}\left(\pi k_{x} \over L_{x}\right)^{2} 
\,B^{(\mathbf{n})}_{\mathbf{k}} , 
\label{Anocoupled}\\
\frac{dB^{(\mathbf{n})}_{\mathbf{k}}}{d\tau} & = & 
\frac{-1}{2\omega_{\mathbf{k}}}\left(\pi k_{x} \over L_{x}\right)^{2} 
\,A^{(\mathbf{n})}_{\mathbf{k}} .
\label{Bnocoupled}
\end{eqnarray}
The solution that satisfies the initial conditions (\ref{iniA}) and 
(\ref{iniB}) reads

\begin{eqnarray}
B^{(\mathbf{n})}_{\mathbf{k}}=\frac{1}{\sqrt{2\omega_{\mathbf{k}}}}\,
\delta_{\mathbf{k},\mathbf{n}}  \cosh(\gamma k_{x} \tau) , 
\label{soluciondiagonal1} \\
A^{(\mathbf{n})}_{\mathbf{k}}=- \frac{1}{\sqrt{2\omega_{\mathbf{k}}}}\,
\delta_{\mathbf{k},\mathbf{n}} \sinh(\gamma k_{x} \tau) ,
\label{soluciondiagonal2}
\end{eqnarray}
where $\gamma=(k_{x}/\Omega) (\pi/L_{x})^2$.
With the help of Eq(\ref{numerodefotones}) we obtain

\begin{equation}
\langle {\mathcal{N}}_{\mathbf{k}} \rangle =   \sinh^{2}(\gamma k_{x} \tau_f) ,
\label{casododonov}
\end{equation}
where $\tau_f=\epsilon t_f$.
In this uncoupled resonance case the average number of created 
photons in the mode ${\mathbf{k}}$ increases exponentially in time 
with a rate given by $2\gamma k_{x}$. 
The same result has been obtained in previous papers 
(cs. Ref.\cite{Dodonov} and Ref.\cite{Plunien}). 
There it was assumed that the coupling condition 
$\omega_{\mathbf{j}}=3\omega_{\mathbf{k}}$ cannot be 
fullfiled for two and three dimensional cavities, essentially due to 
the non-equidistant character of the spectrum. As we shall see, this is 
not always true. In what follows we will solve Eq.(\ref{ec1}) and 
Eq.(\ref{ec2}) with coupled modes, and we will show some explicit examples.

Let us now assume the existence of one mode, say $\mathbf{j}$, in the 
infinite sum in Eq.(\ref{ec1}) and Eq.(\ref{ec2}),  which satisfies 
$\omega_{\mathbf{j}}=3\omega_{\mathbf{k}}$. We obtain for 
$A_{\mathbf{k}}^{(\mathbf{n})}$ and $B_{\mathbf{k}}^{(\mathbf{n})}$ 

\begin{eqnarray}
\frac{dA^{(\mathbf{n})}_{\mathbf{k}}}{d\tau}&=&\gamma\left(-k_{x}
\,B^{(\mathbf{n})}_{\mathbf{k}}+
(-1)^{j_x+k_x}j_{x}\,A^{(\mathbf{n})}_{\mathbf{j}}\right) ,
\label{Akcoupled}\\
\frac{dB^{(\mathbf{n})}_{\mathbf{k}}}{d\tau}&=&\gamma\left(-k_{x}
\,A^{(\mathbf{n})}_{\mathbf{k}}+
(-1)^{j_x+k_x}j_{x}\,B^{(\mathbf{n})}_{\mathbf{j}}\right) ,
\label{Bkcoupled}
\end{eqnarray}
where we have used that the relation 
$\omega_{\mathbf{j}}=3\omega_{\mathbf{k}}$ is equivalent to
\begin{equation}
j_{x}^{2}=9\,k_{x}^{2}+8 \left[ \left( {L_x\over L_y} k_{y}\right)^{2}+
\left( {L_x\over L_z}k_{z} \right)^{2})  \right]
\label{rel}
\end{equation}
because the coupling coefficient $g_{\mathbf{kj}}$ is proportional to 
$\delta_{k_{y}j_{y}}\delta_{k_{z}j_{z}}$.
The next step is to obtain the equations for $A_{\mathbf{j}}^{(\mathbf{n})}$ 
and $B_{\mathbf{j}}^{(\mathbf{n})}$. The mode $\mathbf{j}$ is coupled to 
modes $\mathbf{s}$ that satisfy 
$2\omega_{\mathbf{k}}=|\omega_{\mathbf{j}}\pm\omega_{\mathbf{s}}|$. Since 
$\omega_{\mathbf{j}}=3\,\omega_{\mathbf{k}}$ this relation is satisfied for 
$\omega_{\mathbf{s}}=\omega_{\mathbf{k}}$ (as expected) and for 
$\omega_{\mathbf{s}}=5\,\omega_{\mathbf{k}}$. We assume that the spectrum 
under consideration does not satisfy the latter. 
In this case, the equations read

\begin{eqnarray}
\frac{dA^{(\mathbf{n})}_{\mathbf{j}}}{d\tau}&=& 
- \frac{ (-1)^{j_x+k_x}
\gamma j_{x}}{3} \, A^{(\mathbf{n})}_{\mathbf{k}} ,  
\label{Ajcoupled} \\
\frac{dB^{(\mathbf{n})}_{\mathbf{j}}}{d\tau}&=&
- \frac{ (-1)^{j_x+k_x}\gamma j_{x}}{3}
\,B^{(\mathbf{n})}_{\mathbf{k}}  .
\label{Bjcoupled}
\end{eqnarray}
In order to find the solution to the above equations we write the system in 
matricial form

\begin{equation}
\frac{d\vec{v}}{d\tau}={\mathcal{M}}\,\vec{v} ,
\label{sys}
\end{equation}
where

\begin{eqnarray}
\vec{v}(\tau)=\left(\begin{array}{c}
 B^{(\mathbf{n})}_{\mathbf{k}}(\tau)\\
 A^{(\mathbf{n})}_{\mathbf{k}}(\tau)\\
 B^{(\mathbf{n})}_{\mathbf{j}}(\tau)\\
 A^{(\mathbf{n})}_{\mathbf{j}}(\tau)
\end{array}
\right) \ \ \ ,  \ \ \ \ \ 
\mathcal{M}=\gamma\left(\begin{array}{cccc}
0 & -k_{x} &(-1)^{k_x+j_x} j_{x} & 0 \\
-k_{x} & 0 & 0 & (-1)^{k_x+j_x} j_{x} \\
{-(-1)^{k_x+j_x}j_{x}/3} & 0 & 0 & 0 \\
0 & {-(-1)^{k_x+j_x}j_{x}/3} & 0 & 0 
\end{array}
\right) ,
\end{eqnarray}
and the initial condition reads

\begin{eqnarray}
\vec{v}(0)=\left(\begin{array}{c}
 \frac{1}{\sqrt{2\omega_{\mathbf{k}}}}\,
\delta_{\mathbf{n},\mathbf{k}}\\
 0 \\
 \frac{1}{\sqrt{2\omega_{\mathbf{j}}}}\,
\delta_{\mathbf{n},\mathbf{j}}\\
 0 
\end{array}
\right) .
\label{condicioninicial}
\end{eqnarray}
The solution is easily obtained after diagonalizing $\mathcal{M}$. The 
eigenvalues are given by

\begin{equation}
\lambda=\pm \frac{\gamma k_{x}}{2} \pm \frac{i\gamma}{6} 
\sqrt{|9 k^{2}_{x}-12j^{2}_{x}|} ,
\label{eig}
\end{equation}
where we have used Eq.(\ref{rel}).
The solution can be formally written as 
\begin{equation}
\vec{v}(\tau)={\mathcal{C}}\,e^{{\mathcal{D}}\tau}{\mathcal{C}}^{-1}\,
\vec{v}(0), 
\label{formal}
\end{equation}
where $\mathcal{D}$ is the eigenvalues diagonal matrix and $\mathcal{C}$ is 
the corresponding eigenvectors matrix. 
This means that $A^{(\mathbf{n})}_{\mathbf{k}}$ and 
$A^{(\mathbf{n})}_{\mathbf{j}}$ are linear combinations of exponential 
functions of the eigenvalues in Eq.(\ref{eig}) times $\tau$. The exponential 
growth of $A^{(\mathbf{n})}_{\mathbf{k}}$ and $A^{(\mathbf{n})}_{\mathbf{j}}$ 
is determined by the eigenvalues with positive real part. Looking at 
Eq.(\ref{numerodefotones}) we conclude that the number of created photons, 
in both the mode ${\mathbf{k}}$ and the mode ${\mathbf{j}}$, will increase 
exponentially in time with a rate given by $\gamma k_{x}$.

This is our main result. In the `resonance parametric case'  
the resonant mode may be coupled to some other mode. In this case the 
number of created photons in both modes grows exponentially in time with 
the same rate, which is exactly one half of the rate expected for 
the resonant mode when the coupling is neglected.

We have derived Eq.(\ref{sys}) assuming that only two modes are coupled. If 
the spectrum contains one mode $\mathbf{s}$ such that 
$\omega_{\mathbf{s}}=5\omega_{\mathbf{k}}$, besides 
$\omega_{\mathbf{j}}=3\omega_{\mathbf{k}}$, we get three coupled modes. The 
resulting equation will be similar to Eq.(\ref{sys}) but with a 
$6\times6$ matrix to diagonalize. The number of photons in each mode 
(${\mathbf{s}}, \ {\mathbf{j}} \ \mathrm{and} \ {\mathbf{k}}$) will grow 
exponentially in time. Due to the non-equidistant character of the
spectrum, it is not common to have three modes coupled. For a cubic 
cavity (i.e. $L_{x}= L_{y}= L_{z} = L$), the
first three modes coupled are   ${\mathbf{k}}=(11,16,13), 
{\mathbf{j}}=(67,16,13),$ and ${\mathbf{s}}=(115,16,13)$, the 
frequency of the lowest mode being an order of magnitude 
larger than the fundamental
frequency of the cavity. This case is therefore of less
interest.

Let us now discuss briefly what happens with the remaning cases in which the 
MSA is non trivial and differs from the naive perturbation approach. We first 
study two non-resonant modes $\mathbf{s}$ and $\mathbf{p}$ (i.e. 
$\omega_{\mathbf{s}}\not= \Omega/2 \not=\omega_{\mathbf{p}}$) satisfying 
the condition (\ref{cond1}), $\omega_{\mathbf{s}}+ \omega_{\mathbf{p}}=\Omega$.
\footnote{Note that we are not necessarily within 
the `parametric resonance case'. 
The external frequency $\Omega$ could be or not twice the 
frequency of some other unperturbed mode.}.
After some algebra on Eqs.(\ref{ec1}) and (\ref{ec2}) we get for 
$A^{(\mathbf{n})}_{\mathbf{s}}$ and $B^{(\mathbf{n})}_{\mathbf{p}}$ the 
following

\begin{eqnarray}
\frac{dA^{(\mathbf{n})}_{\mathbf{s}}}{d\tau}=-\frac{1}{2
\omega_{\mathbf{s}}}\left({\pi\over L_{x}}\right)^{2} (-1)^ {s_{x}+p_{x}} 
s_{x}p_{x} B^{(\mathbf{n})}_{\mathbf{p}} , \\
\frac{dB^{(\mathbf{n})}_{\mathbf{p}}}{d\tau}=-\frac{1}{2\omega_{\mathbf{p}}}
\left({\pi\over L_{x}}\right)^{2} (-1)^ {s_{x}+p_{x}} s_{x}p_{x}
A^{(\mathbf{n})}_{\mathbf{s}} ,
\end{eqnarray}
and same equations holding for $B^{(\mathbf{n})}_{\mathbf{s}}$ 
and $A^{(\mathbf{n})}_{\mathbf{p}}$.
The solutions are straightforwardly obtained,
giving for the average value of the number operator 
\begin{equation}
\langle{\mathcal{N}}_{\mathbf{s}} \rangle=\langle{\mathcal{N}}_
{\mathbf{p}}\rangle=\sinh^{2}\left(\left({\pi\over L_{x}}
\right)^{2}\,\frac{s_{x}p_{x}}{2\sqrt{\omega_{\mathbf{s}}
\omega_{\mathbf{p}}}}\,{\tau_f}\right) .
\end{equation}
Note that if we set $s_{x}=p_{x}\,(\mathbf{s}=\mathbf{p})$ we recover 
the `parametric resonance case'. This example
shows the possibility of obtaining exponential growth of photons in modes 
which are not in resonance with the external frequency.

If the spectrum contains some sequence of equidistant frequencies 
$\omega_{{\mathbf{p}}_{i}}$ separated by $\Omega$, the corresponding 
modes will be coupled through the conditions 
$\Omega=\omega_{{\mathbf{p}}_{i+1}} - \omega_{{\mathbf{p}}_{i}}$ 
and $\Omega=\omega_{{\mathbf{p}}_{i}} - \omega_{{\mathbf{p}}_{i-1}}$. 
One can show that, as long as the modes ${\mathbf{p}}_{i}$ are not coupled to 
modes outside the sequence, the number of created photons in each of 
these modes will be an oscillatory function of time.

\subsection{Examples}
\label{examples}

The first and more important example is the cubic  cavity. 
In order to obtain 
`parametric resonance' we fix $\Omega$ as twice the lowest cavity 
frequency,

\begin{equation}
\Omega=2\omega_{(1,1,1)} = \frac{2 \pi \sqrt{3}}{L} .
\label{omega1}
\end{equation}
For this example we will assume that $L=1cm$.
The fundamental mode ${\mathbf{k}}=(1,1,1)$ will be coupled to 
${\mathbf{j}}=(5,1,1)$ because $\omega_{(5,1,1)}=3\omega_{(1,1,1)}$. 
Only these two modes are coupled, since there does not exist in the spectrum 
any mode ${\mathbf{s}}$ satisfaying $\omega_{\mathbf{s}}=5\omega_{(1,1,1)}$.
The exponential growth for the modes ${\mathbf{k}}$ and ${\mathbf{j}}$
will be one half of that expected by previous 
papers \cite{Plunien}. Now we can write explicity 
Eq.(\ref{formal}) for this particular case. The result 
is

\begin{eqnarray}
B^{(\mathbf{n})}_{\mathbf{k}}(\tau)&=&
\frac{ \delta_{\mathbf{n},\mathbf{k}}}{\sqrt{2\omega_{\mathbf{k}}}}
[\cos(2.56\tau)\cosh(0.45\tau)+0.176\sin
(2.57\tau)\sinh(0,45\tau)] \nonumber \\                                       
                                     & &\ \ \ \ \ \ \ \ \ \ +
\frac{ \delta_{\mathbf{n},\mathbf{j}}}{\sqrt{2\omega_{\mathbf{j}}}}[1.76
\sin(2.57\tau)\cosh(0.45\tau)] , \label{uno}\\
A^{(\mathbf{n})}_{\mathbf{k}}(\tau)&=&
\frac{\delta_{\mathbf{n},\mathbf{k}}}{\sqrt{2\omega_{\mathbf{k}}}}
[\cos(2.56\tau)\sinh(0.45\tau)
+0.176\sin(2.57\tau)\cosh(0.45\tau)] \nonumber \\
                                     & &\ \ \ \ \ \ \ \ \ \ 
+\frac{\delta_{\mathbf{n},\mathbf{j}}}{\sqrt{2\omega_{\mathbf{j}}}}
[1.76\sin(2.57\tau)\sinh(0..45\tau)] , \label{a1}\\
B^{(\mathbf{n})}_{\mathbf{j}}(\tau)&=&
\frac{\delta_{\mathbf{n},\mathbf{k}}}{\sqrt{2\omega_{\mathbf{k}}}}
[-0.586\sin(2.56\tau)\cosh(0.45\tau)] \nonumber \\
                                     & &\ \ \ \ \ \ \ \ \ \ 
+\frac{\delta_{\mathbf{n},\mathbf{j}}}{\sqrt{2\omega_{\mathbf{j}}}}
[\cos(2.56\tau)\cosh(0.45\tau)-0.176\sin(2.56\tau)\sinh(0.45\tau)] , 
\label{dos} \\   
A^{(\mathbf{n})}_{\mathbf{j}}(\tau)&=&
\frac{\delta_{\mathbf{n},\mathbf{k}}}{\sqrt{2\omega_{\mathbf{k}}}}
[-0.586\sin(2.56\tau)\sinh(0.45\tau)] \nonumber \\
                                     & &\ \ \ \ \ \ \ \ \ \ 
+\frac{\delta_{\mathbf{n},\mathbf{j}}}{\sqrt{2\omega_{\mathbf{j}}}}
[\cos(2.56\tau)\sinh(0.45\tau)-0.176\sin(2.56\tau)\cosh(0.45\tau)] . 
\label{a2} 
\end{eqnarray}
An important remark is that this solution satisfies the unitary condition for 
the Bogoliubov transformation (\ref{bog1}),
\begin{equation}
\sum_{\mathbf{n}}\, |B^{(\mathbf{n})}_{\mathbf{k}}|^{2}-
|A^{(\mathbf{n})}_{\mathbf{k}}|^{2}=\frac{1}{\sqrt{2\omega_{\mathbf{k}}}} .
\end{equation}
We can compute the number of created photons  
in each mode inserting Eqs.(\ref{a1}) and (\ref{a2}) 
into Eq.(\ref{numerodefotones}). The result is

\begin{eqnarray}
\langle{\mathcal{N}}_{\mathbf{k}}\rangle=\cos^{2}(2.56\tau_{f})
\sinh^{2}(0.45\tau_{f})+1.06\sin^{2}(2.56\tau_{f})
\cosh^{2}(0.45\tau_{f}) + 0.088\sin(5.12\tau_{f}) \sinh(0.9\tau_{f}) , \\
\langle{\mathcal{N}}_{\mathbf{j}}\rangle=\cos^{2}(2.56\tau_{f})
\sinh^{2}(0.45\tau_{f})+1.06\sin^{2}(2.56\tau_{f})\cosh^{2}(0.45\tau_{f})
-0.088\sin(5.12\tau_{f})\sinh(0.9\tau_{f}) .
\end{eqnarray}
When $\tau_{f}\geq 1$ these expressions are approximated by

\begin{equation}
\langle{\mathcal{N}}_{\mathbf{k}}
\rangle\approx\langle{\mathcal{N}}_{\mathbf{j}}\rangle \approx 
e^{0.9\tau_{f}} .
\label{approx}
\end{equation}

In a previous paper \cite{Dodonov} the authors considered 
two dimensional cavities, which means that one of the cavity's 
dimensions is much smaller than the others (say $L_{z}<<L_{x},L_{y}$). 
We can easily recover this limit by omitting the `z' 
dimension. In what follows we will discuss this case, for 
increasing external frequencies.

Let us first assume that $L_{x}=L_{y}$. 
If $\Omega=2\omega_{(1,1)}$, then the fundamental mode ${\bf k}=(1,1)$ does 
not couple to any other mode and it grows exponentially in time.
The next resonant frequencies are $\Omega=2\omega_{(1,2)}$ 
and $\Omega=2\omega_{(2,2)}$. In both cases the resonant mode is 
not coupled. If $\Omega=2\omega_{(1,3)}$, the mode $(1,3)$ will be coupled 
to the 
mode $(9,3)$ and both will grow exponentially in time. However, the 
mode $(3,1)$ also satisfies the `parametric resonance condition' and,
being uncoupled to other modes, it will grow faster than the previous ones.
For the same frequency, $\Omega=2\omega_{(1,3)}$, we have found
by inspection 
three equidistant modes, $\omega_{(13,39)}=13\pi\sqrt{10}/L$, 
$\omega_{(27,39)}=15\pi\sqrt{10}/L$ and 
$\omega_{(37,39)}=17\pi\sqrt{10}/L$. 
The number of created photons in each mode will oscillate in time.

Now we choose $L_{x}=3L_{y}$. This choice makes the fundamental mode to be 
coupled in `parametric resonance'. 
If we set $\Omega=2\omega_{(1,1)}$, the mode $(9,1)$ satisfies that 
$\omega_{(9,1)}=3\omega_{(1,1)}$, so both modes will grow exponentially.

\subsection{Off resonance}
\label{offresonance}
 
In this subsection we study what happens when the external frequency 
$\tilde{\Omega}$ is slightly off-resonance, i.e., $\tilde{\Omega} = 
\Omega + h$, where $\Omega$ is a resonant frequency and $h \ll \Omega$.
We will assume that $h = \epsilon\alpha$, where
$\alpha = O(\Omega)$.
We will show how to apply MSA to this case. Off-resonance motions have
already been considered in the literature \cite{Dod3} using
a different approach, and it was shown that there are threshold conditions
on $h$ for exponential photon creation.

For small amplitudes of oscillations, the equations for the modes are still
Eq.(\ref{eqqk}) with $\Omega$  replaced by the external frequency 
$\tilde{\Omega}$. Since $h \ll \Omega$, all factors of the form
$e^{\pm i h t}$ may be regarded as slow oscillations, so that the MSA
conditions to get rid of secularities are exactly the same as in the resonant
case, Eqs.(\ref{cond1},\ref{cond2},\ref{cond3}). However, the equations
for the modes (\ref{ec1},\ref{ec2}) do get modified. Their off-resonant 
version reads

\begin{eqnarray}
{d A^{({\bf n})}_{\bf k}\over d \tau} &=&
-\frac{\pi^2 k_x^2}{2 \omega_{\bf k} L_x^2} e^{i \alpha \tau} 
B^{({\bf n})}_{\bf k}
\delta(2 \omega_{\bf k} -\Omega) +
\sum_{\bf j} (-\omega_{\bf j} + \frac{\Omega+h}{2})
\delta(-\omega_{\bf k} - \omega_{\bf j} + \Omega) 
\frac{\Omega+h}{2 \omega_{\bf k}} g_{{\bf k}{\bf j}} e^{i \alpha \tau}
B^{({\bf n})}_{\bf j} \nonumber \\
&& + \sum_{\bf j} \left[
e^{i \alpha \tau} (\omega_{\bf j} + \frac{\Omega+h}{2}) 
\delta(\omega_{\bf k} - \omega_{\bf j} - \Omega) +
e^{-i \alpha \tau} (\omega_{\bf j} - \frac{\Omega+h}{2}) 
\delta(\omega_{\bf k} - \omega_{\bf j} + \Omega)
\right]
\frac{\Omega+h}{2 \omega_{\bf k}} g_{{\bf k}{\bf j}} A^{({\bf n})}_{\bf j} ,
\label{offec1}
\end{eqnarray}
and

\begin{eqnarray}
{d B^{({\bf n})}_{\bf k}\over d\tau} &=&
-\frac{\pi^2 k_x^2}{2 \omega_{\bf k} L_x^2} e^{-i \alpha \tau} 
A^{({\bf n})}_{\bf k}
\delta(2 \omega_{\bf k} -\Omega) +
\sum_{\bf j} (-\omega_{\bf j} + \frac{\Omega+h}{2})
\delta(-\omega_{\bf k} - \omega_{\bf j} + \Omega) 
\frac{\Omega+h}{2 \omega_{\bf k}} g_{{\bf k}{\bf j}} 
e^{-i \alpha \tau} A^{({\bf n})}_{\bf j} \nonumber \\
&& + \sum_{\bf j} 
\left[
e^{-i \alpha \tau} (\omega_{\bf j} + \frac{\Omega+h}{2}) 
\delta(\omega_{\bf k} - \omega_{\bf j} - \Omega) +
e^{i \alpha \tau} (\omega_{\bf j} - \frac{\Omega+h}{2}) 
\delta(\omega_{\bf k} - \omega_{\bf j} + \Omega)
\right]
\frac{\Omega+h}{2 \omega_{\bf k}} g_{{\bf k}{\bf j}} B^{({\bf n})}_{\bf j} .
\label{offec2}
\end{eqnarray} 

Let us solve these equations in the quasi parametric resonant case, that is
for $\tilde{\Omega}- h = \Omega = 2 \omega_{\bf k}$. In the case when there
are no coupled modes, we get two coupled first order differential equations
for $A^{(\mathbf{n})}_{\mathbf{k}}$ and $B^{(\mathbf{n})}_{\mathbf{k}}$.
After the change of variables
\begin{eqnarray}
A^{(\mathbf{n})}_{\mathbf{k}} &=& e^{i \alpha \tau/2} 
a^{(\mathbf{n})}_{\mathbf{k}} , \nonumber\\
B^{(\mathbf{n})}_{\mathbf{k}} &=& e^{-i \alpha \tau/2} 
b^{(\mathbf{n})}_{\mathbf{k}} ,
\label{change}
\end{eqnarray}
the equations take the form
\begin{eqnarray}
\frac{da^{(\mathbf{n})}_{\mathbf{k}}}{d\tau} & = & 
- i \frac{\alpha}{2} a^{(\mathbf{n})}_{\mathbf{k}} - \gamma k_x
b^{(\mathbf{n})}_{\mathbf{k}} , \\
\frac{db^{(\mathbf{n})}_{\mathbf{k}}}{d\tau} & = & 
i \frac{\alpha}{2} b^{(\mathbf{n})}_{\mathbf{k}} - \gamma k_x
a^{(\mathbf{n})}_{\mathbf{k}} .
\end{eqnarray}

There will be
growing exponential solutions if an eigenvalue of the corresponding
matrix 
\begin{eqnarray}
\left (\begin{array}{cc}
-{i\alpha\over 2} & -\gamma k_x\\
-\gamma k_x & {i\alpha\over 2} 
\end{array}
\right)
\end{eqnarray}
has a positive real part. The eigenvalues are given by 
\begin{equation}
\lambda = \pm\sqrt{\gamma^2 k_x^2 - {\alpha^2\over 4}}
\end{equation}
and lead to the 
following threshold for resonant behavior
\begin{equation}
|h| < 2 \epsilon \gamma k_x \iff {|h|\over \Omega}< {\epsilon\over 2}
{\left (\frac{k_{x}}{L_{x}}\right)^{2}\over 
\left(\frac{k_{x}}{L_{x}}\right)^{2} + 
\left(\frac{k_{y}}{L_{y}}\right)^{2} + 
\left(\frac{k_{z}}{L_{z}}\right)^{2}} .
\label{thr1}
\end{equation} 
For this uncoupled, quasi-resonant situation, the resonant mode 
satisfies a Mathieu equation (see Eq.(\ref{eqqk})). The threshold
we obtained in Eq. (\ref{thr1}) coincides with the one 
obtained for the Mathieu equation using a different method
\cite{Landau}.
 
As a second example, let us consider the case of two coupled modes, say
modes ${\bf k}$ and ${\bf j}$, for which $\omega_{\bf j}=3 \omega_{\bf k}$.
After the change of variables (\ref{change}), 
and defining
\begin{eqnarray}
A^{(\mathbf{n})}_{\mathbf{j}} &=& e^{3 i \alpha \tau/2} 
a^{(\mathbf{n})}_{\mathbf{j}} , \nonumber\\
B^{(\mathbf{n})}_{\mathbf{j}} &=& e^{- 3 i \alpha \tau/2} 
b^{(\mathbf{n})}_{\mathbf{j}} ,
\label{change2}
\end{eqnarray}
the off-resonance form of 
Eqs.(\ref{Akcoupled},\ref{Bkcoupled},\ref{Ajcoupled},\ref{Bjcoupled}) is

\begin{eqnarray}
\frac{da^{(\mathbf{n})}_{\mathbf{k}}}{d\tau}&=&
-i \frac{\alpha}{2} a^{(\mathbf{n})}_{\mathbf{k}} +
\gamma \left( -k_{x}
\,b^{(\mathbf{n})}_{\mathbf{k}}+
(-1)^{j_x+k_x}j_{x}  
\, a^{(\mathbf{n})}_{\mathbf{j}}\right) , \\
\frac{db^{(\mathbf{n})}_{\mathbf{k}}}{d\tau}&=&
i \frac{\alpha}{2} b^{(\mathbf{n})}_{\mathbf{k}} +
\gamma\left(-k_{x}
\,a^{(\mathbf{n})}_{\mathbf{k}}+
(-1)^{j_x+k_x}j_{x}   
\, b^{(\mathbf{n})}_{\mathbf{j}}\right) , \\
\frac{da^{(\mathbf{n})}_{\mathbf{j}}}{d\tau}&=& 
-3 i \frac{\alpha}{2} a^{(\mathbf{n})}_{\mathbf{j}}
- (-1)^{j_x+k_x} {\gamma j_{x}\over 3}   
\, a^{(\mathbf{n})}_{\mathbf{k}} ,  \\
\frac{db^{(\mathbf{n})}_{\mathbf{j}}}{d\tau}&=&
 3 i \frac{\alpha}{2} b^{(\mathbf{n})}_{\mathbf{j}}
- (-1)^{j_x+k_x} {\gamma j_{x}\over 3} 
\, b^{(\mathbf{n})}_{\mathbf{k}} ,
\end{eqnarray}
where we have neglected terms proportional to $h/\omega_{\bf k}
=O(\epsilon)$ in the r.h.s., since they would introduce only small
corrections to the eigenvalues and thresholds.
Just as in subsection A, we can rewrite these 
equations in matricial form, and 
find the solution by diagonalizing such a matrix. The corresponding eigenvalues
are

\begin{equation}
\lambda = \pm \frac{1}{\sqrt{2}} \sqrt{ - U \pm \sqrt{V}} ,
\end{equation}
where 
$ U= \frac{5 \alpha^2}{2} + \frac{\gamma^2}{3} [ 2 j_x^2 - 3 k_x^2 ] $ 
is a positive
number (see Eq. (\ref{rel})), and

\begin{equation}
V = 4 \alpha^4 + 
4 \alpha^2 \gamma^2  ( k_x^2 + {4\over 3} j_x^2 ) +
\gamma^4 k_x^2  ( k_x^2 - {4\over 3} j_x^2) \,\, .
\end{equation}
For $\alpha=0$, $V$ is negative, so there are two eigenvalues with positive
real part, which corresponds to exponential growth. As $|\alpha|$ grows,
$V$ will eventually become positive, all eigenvalues are pure imaginary
for $\sqrt {V}<U$,
and no exponential growth is obtained. The condition
for exponential growth is $V<0$, which again sets a threshold for 
the off-resonance frequency difference $h$,

\begin{equation}
| h | < \epsilon \gamma \sqrt{-d_1 + \sqrt{d_1^2 + d_2}} \,\, ,
\end{equation}
with
\begin{eqnarray}
d_1 &=&  ( k_x^2 + \frac{4}{3} j_x^2) > 0 , \\
d_2 &=& k_x^2 (\frac{4}{3} j_x^2 - k_x^2) > 0 .
\end{eqnarray}
For the example of the cubic cavity discussed in the previous subsection,
the threshold is $|h|< 0.68\epsilon\gamma$, that is
$|h|/\Omega < 0.06\epsilon$.

\section{Dynamical casimir effect at finite temperature}

Up to this point we have assumed that the field was in the in-vacuum state 
$T=0$ (see, for example, Eq.(\ref{numerodefotones})). It is well known
that the temperature contribution can dominate the pure vacuum effect when 
computing the {\textit{static}} Casimir force \cite{review}. 
Then, it is expected that temperature plays an important role in the 
dynamical Casimir effect as well.
In what follows we shall derive an expression for the number of created 
photons inside the cavity equivalent to Eq.(\ref{numerodefotones}),  but now 
assuming an initial state in equilibrium at finite temperature. After that, 
we will apply the obtained result to the resonant vibrating cavity.

We adopt the scheme of scalar field quantization developed in section II. 
For $t<0$ the wall is at rest, so we assume the system to be at thermal 
equilibrium at finite temperature $T=1/\beta$. The state of the 
system is described by a statistical operator $\rho$ which does not evolve 
in time. We expand this operator in Fock states of the field at $t<0$ 
\begin{eqnarray}
\rho=\frac{1}{Z}\sum_{n_{{\mathbf{k}}_{1}}\geq0}
\sum_{n_{{\mathbf{k}}_{2}}\geq0} \ldots e^{-\beta\sum_{i}\,
(n_{{\mathbf{k}}_{i}}+{1\over2})E_{{\mathbf{k}}_{i}}}|n_{{\mathbf{k}}_{1}} 
n_{{\mathbf{k}}_{2}} \ldots \rangle \langle n_{{\mathbf{k}}_{1}}
n_{{\mathbf{k}}_{2}} \ldots | ,
\label{rho}
\end{eqnarray}
where $E_{{\mathbf{k}}_{i}}=\omega_{{\mathbf{k}}_{i}}$ and the 
normalization factor is given by

\begin{equation}
Z=\sum_{n_{{\mathbf{k}}_{1}}\geq0}\sum_{n_{{\mathbf{k}}_{2}}\geq0}
\ldots e^{-\beta\sum_{i}\,(n_{{\mathbf{k}}_{i}}+{1\over2})
E_{{\mathbf{k}}_{i}}} .
\label{zeta}
\end{equation}

At $t=0$ the wall begins to oscillate and the system will no longer remain 
at thermal equilibrium. Following the steps given in section II we assume 
that the wall stops at $t=t_{\rm final}$. In that case, the number of photons 
in the $\mathbf{k}$-mode is the average value of 
$\hat{a}_{{\mathbf{k}}}^{\dag\,\scriptscriptstyle{\rm out}}
\hat{a}_{{\mathbf{k}}}^{\scriptscriptstyle{\rm out}}$ with respect to the 
initial state $\rho$

\begin{equation}
\langle{\mathcal{N}}_{\mathbf{k}}\rangle_{\rho}=
{\rm Tr}(\rho\hat{a}_{{\mathbf{k}}}^{\dag\,\scriptscriptstyle{\rm out}}
\hat{a}_{{\mathbf{k}}}^{\scriptscriptstyle{\rm out}}).
\label{trasa}
\end{equation}
Using Eq.(\ref{bog1}) we can write this expression in terms of 
$\hat{a}^{\scriptscriptstyle{\rm in}}$ and 
$\hat{a}^{\dag\,\scriptscriptstyle{\rm in}}$

\begin{eqnarray}
{\rm Tr}(\rho\,\hat{a}_{{\mathbf{k}}}^{\dag\,\scriptscriptstyle{\rm out}}
\hat{a}_{{\mathbf{k}}}^{\scriptscriptstyle{\rm out}})=
\sum_{\mathbf{n\,j}}\beta_{\mathbf{n\,k}}\,\alpha_{\mathbf{j\,k}}\,
{\rm Tr}(\rho\,\hat{a}_{{\mathbf{n}}}^{\scriptscriptstyle{\rm in}}
\hat{a}_{{\mathbf{j}}}^{\scriptscriptstyle{\rm in}})+
\beta_{\mathbf{n\,k}}\,\beta_{\mathbf{j\,k}}^{\star}\,
{\rm Tr}(\rho\,\hat{a}_{{\mathbf{n}}}^{\scriptscriptstyle{\rm in}}
\hat{a}_{{\mathbf{j}}}^{\dag\,\scriptscriptstyle{\rm in}})
\nonumber \\
+\alpha^{\star}_{\mathbf{n\,k}}\,\alpha_{\mathbf{j\,k}}^{\star}\,
{\rm Tr}(\rho\,\hat{a}_{{\mathbf{n}}}^{\dag\,\scriptscriptstyle{\rm in}}
\hat{a}_{{\mathbf{j}}}^{\scriptscriptstyle{\rm in}})+
\alpha_{\mathbf{n\,k}}^{\star}\,\beta_{\mathbf{j\,k}}^{\star}\,
{\rm Tr}(\rho\,\hat{a}_{{\mathbf{n}}}^{\dag\,\scriptscriptstyle{\rm in}}
\hat{a}_{{\mathbf{j}}}^{\dag\,\scriptscriptstyle{\rm in}}).
\label{desarrollo}
\end{eqnarray}
With the help of Eqs.(\ref{rho}) and (\ref{zeta}) it is a 
straightforward calculation to find that

\begin{eqnarray}
{\rm Tr}(\rho\,\hat{a}_{{\mathbf{n}}}^{\scriptscriptstyle{\rm in}}
\hat{a}_{{\mathbf{j}}}^{\scriptscriptstyle{\rm in}})&=&0=
{\rm Tr}(\rho\,\hat{a}_{{\mathbf{n}}}^{\dag\,\scriptscriptstyle{\rm in}}
\hat{a}_{{\mathbf{j}}}^{\dag\,\scriptscriptstyle{\rm in}})  \\
{\rm Tr}(\rho\,\hat{a}_{{\mathbf{n}}}^{\dag\,\scriptscriptstyle{\rm in}}
\hat{a}_{{\mathbf{j}}}^{\scriptscriptstyle{\rm in}})&=&
\frac{1}{e^{\beta E_{\mathbf{n}}}-1}\,\delta_{\mathbf{n\,j}}=
{\rm Tr}(\rho\,\hat{a}_{{\mathbf{n}}}^{\scriptscriptstyle{\rm in}}
\hat{a}_{{\mathbf{j}}}^{\dag\,\scriptscriptstyle{\rm in}})-1 .
\end{eqnarray} 
Inserting this into Eqs.(\ref{desarrollo}) and (\ref{trasa}) we arrive at

\begin{equation}
\langle{\mathcal{N}}_{\mathbf{k}}\rangle_{\rho}=
\sum_{\mathbf{n}}|\beta_{\mathbf{n\,k}}|^{2}+
\sum_{\mathbf{n}}(|\beta_{\mathbf{n\,k}}|^{2}+
|\alpha_{\mathbf{n\,k}}|^{2})\,\frac{1}{e^{\beta E_{\mathbf{n}}}-1} .
\end{equation}
Finally, using Eqs.(\ref{bog2}) and (\ref{numerodefotones}) we get
\begin{equation}
\langle{\mathcal{N}}_{\mathbf{k}}\rangle_{\rho}=
\langle{\mathcal{N}}_{\mathbf{k}}\rangle_{T=0}+
\sum_{\mathbf{n}}(|B^{(\mathbf{n})}_{\mathbf{k}}|^{2}+
|A^{(\mathbf{n})}_{\mathbf{k}}|^{2}) 
\frac{2\omega_{\mathbf{k}}}{e^{\beta E_{\mathbf{n}}}-1} .
\label{final}
\end{equation}

Let us now apply the results obtained in section IV for the vibrating cavity 
to the case in which the field is initially at thermal equilibrium. In the 
parametric resonance case without coupling the Bogoliubov coefficients are 
diagonal (see Eqs.(\ref{soluciondiagonal1}) and (\ref{soluciondiagonal2})), 
so Eq.(\ref{final}) can be reduced to

\begin{equation}
\langle{\mathcal{N}}_{\mathbf{k}}\rangle_{\rho}=
\langle{\mathcal{N}}_{\mathbf{k}}\rangle_{T=0}
\left( 1+2\frac{1}{e^{\beta E_{\mathbf{k}}}-1} \right) +
\frac{1}{e^{\beta E_{\mathbf{k}}}-1} ,
\label{Plunien}
\end{equation}
with $\langle{\mathcal{N}}_{\mathbf{k}}\rangle_{T=0}$ given 
by Eq.(\ref{casododonov}).
We see that the effect of the temperature is to 
enhance the amount of created photons in the pure vacuum case by a thermal 
distribution factor. Note that the second term in expression (\ref{Plunien}) 
corresponds to the average number of photons, in the mode $\mathbf{k}$, 
present in the cavity when $t<0$.
The same result was obtained in Ref.\cite{Plunien} for the fundamental mode 
of a cubic cavity using  a different approach. However, we have seen that in 
this case the Bogoliubov coefficients are not diagonal, due to the coupling 
between the fundamental mode ${\mathbf{k}}=(1,1,1)$ and the mode 
${\mathbf{j}}=(5,1,1)$. With the help of Eqs.(\ref{uno}), (\ref{a1}), 
(\ref{dos}), and (\ref{a2}) it is easy to obtain the number of photons 
present in the fundamental mode after the wall stops
\begin{eqnarray}
\langle{\mathcal{N}}_{\mathbf{k}}\rangle_{\rho}=
\langle{\mathcal{N}}_{\mathbf{k}}\rangle_{T=0}
(1+2n_{\mathbf{k}})-2\sin^{2}(2.56\tau)\cosh^{2}(0.45\tau)
(n_{\mathbf{k}}-n_{\mathbf{j}}) \nonumber \\ 
-\sin^{2}(2,56\tau)(n_{\mathbf{k}}+n_{\mathbf{j}})+ n_{\mathbf{k}} ,   
\end{eqnarray}
where $n_{\mathbf{k}}=1/(e^{\beta E_{\mathbf{k}}}-1)$ is the Bose mean 
ocupation number for $t<0$.
For the mode $\mathbf{j}$ we find
\begin{eqnarray}
\langle{\mathcal{N}}_{\mathbf{j}}\rangle_{\rho}=
\langle{\mathcal{N}}_{\mathbf{j}}\rangle_{T=0}
(1+2n_{\mathbf{j}})-2\sin^{2}(2.56\tau)\cosh^{2}(0.45\tau)
(n_{\mathbf{j}}-n_{\mathbf{k}}) \nonumber \\ 
-\sin^{2}(2.56\tau)(n_{\mathbf{j}}+n_{\mathbf{k}})+ n_{\mathbf{j}} .
\end{eqnarray}
Again, we obtain that the effect of the temperature is to increase the pure 
vacuum case with thermal factors now depending on $t$.
For $\tau_{f}\geq1$ we have $\langle{\mathcal{N}}_{\mathbf{k}}\rangle_{T=0} 
\approx\langle{\mathcal{N}}_{\mathbf{j}}\rangle_{T=0}$
(see Eq. \ref{approx}). Therefore the total number of photons created inside 
the cavity becomes

\begin{equation}
\langle{\mathcal{N}}_{\rm total}\rangle_{\rho}\cong
(1+n_{\mathbf{k}}+n_{\mathbf{j}}) 
\langle{\mathcal{N}}_{\rm total}\rangle_{T=0} .
\end{equation}

For a cubic cavity of size $L=1 {\rm cm}$ 
at room temperature $T\approx 290{\rm K}$, 
we obtain that the total number of created photons is approximately 300 times 
that of pure vacuum ($T=0$).


\section{CONCLUSIONS}

We have calculated the photon production inside a three 
dimensional
oscillating cavity,
using MSA to deal with
the resonant effects. We have taken into account that, even though
the spectrum of the cavity is non equidistant, 
the different modes may be coupled, and  this coupling 
affects the rate of photon creation.

We have found resonant effects when the external frequency is 
equal to the sum of the frequencies of
two unperturbed modes 
$\Omega=\omega_{\mathbf{s}}+ \omega_{\mathbf{p}}$. When 
$\omega_{\mathbf{s}}\neq \omega_{\mathbf{p}}$, the number 
of photons in both modes grows exponentially. When
$\Omega = 2 \omega_{\mathbf{k}}$, the usual ``parametric resonance case'',
the number of photons in the mode ${\mathbf k}$ also 
grows exponentially, along with the number of photons of 
other modes coupled to ${\mathbf k}$. When the mode  ${\mathbf k}$ 
is coupled to one mode,
the rate 
of photon creation decreases by a
factor 2 with respect to the uncoupled case.

We have also analyzed slightly off-resonance situations. Using an
extension of the MSA we showed that the number of photons in the
relevant modes also increases exponentially if certain threshold
conditions are satisfied. These conditions imply that the external
frequency should be almost equal to the resonant frequency.

As an important example, we have described in detail the case 
of a cubic cavity. The fundamental mode  $(1,1,1)$
is coupled to the mode $(5,1,1)$ when the external frequency is 
$\Omega=2\omega_{(1,1,1)}$.  The number of photons
created in both modes grows as $e^{0.9\tau}$. Neglecting
the mode coupling, one would erroneously conclude that
the number of photons in the mode $(1,1,1)$ grows as
$e^{1.8\tau}$, and that there is no exponential growth
for the mode $(5,1,1)$.
The mode coupling in three dimensional cavities has not been taken 
into account in the previous literature.

We have also computed the enhancement of the dynamical Casimir
effect for an  initial thermal state. The main result is
contained in Eq. (\ref{final}). Only when the Bogoliubov
coefficients are approximately diagonal one recovers the 
usual result, i.e. the number of photons created in a given mode
at temperature $T$ is the $T=0$ result times the thermal
distribution factor. 

For simplicity, we have considered a quantum scalar field. The 
generalization to the case of an electromagnetic field is not
completely straightforward. We state here the main results,
the details will be described in a future work.
Assuming that the potential vector satisfies the gauge 
condition $\nabla . {\mathbf A} =0$, the Maxwell
equations read $\Box{\mathbf A}=0$.
The boundary conditions are the usual (perfect conductors) on the
static walls. On the moving mirror, these boundary conditions
must be imposed in a Lorentz frame in which the mirror is 
instantaneously at rest (see \cite{Neto}).
This implies that, at $x = L_x(t)$, the potential vector must
satisfy $A_y(t, L_x(t),y,z)=A_z(t, L_x(t),y,z)=0$.

As for the case of the scalar field, one can expand the potential
vector in an instantaneous basis, and use MSA to deal with the 
secular terms. The resonant conditions coincide with  Eqs. \ref
{cond1} - \ref{cond3}
in the present paper. The differential equations for the
(now polarization dependent) Bogoliubov coefficients are 
different for the transverse electric TE and transverse magnetic TM 
cases. However, it can be shown that, in the parametric resonant
case, the rate of photon production in both polarizations
TE and TM is the same and coincides with the rate 
for the scalar field computed in this paper.

Finally, we would like to comment briefly  about the possibility of
observing experimentally the dynamical Casimir effect.
Considering cavities of dimensions of the order of $1 {\rm cm}$, the external
frequency should be at least $1 {\rm GHz}$ in order to have resonant
photon creation. This is not trivial, the upper
limit being hopefully around $100 {\rm MHz}$ with the present 
techniques \cite{dono}. Other serious technical problem is that, as
already mentioned, in order
to have resonance the external frequency should be tuned with 
high accuracy to the resonant frequency. Although
extremely difficult, an experimental verification 
of the photon production seems  not completely unrealistic.


\section{ACKNOWLEDGEMENTS}

We are grateful to Paulo A. Maia Neto for reading the manuscript. D.D. thanks 
Luis M. Bettencourt for discussions. The work of M.C. and F.D.M.
was supported by Universidad de Buenos Aires, Conicet, and Agencia 
Nacional de Promoci\'on Cient\'\i fica y Tecnol\'ogica, Argentina. 


\end{document}